\documentstyle[12pt,epsfig]{article}  
\begin{document} 
\baselineskip=20pt

\def\la{\mathrel{\mathpalette\fun <}}
\def\ga{\mathrel{\mathpalette\fun >}}
\def\fun#1#2{\lower3.6pt\vbox{\baselineskip0pt\lineskip.9pt
\ialign{$\mathsurround=0pt#1\hfil##\hfil$\crcr#2\crcr\sim\crcr}}} 

\begin{titlepage} 
\begin{center}
{\Large \bf Extracting the jet azimuthal anisotropy from 
higher order cumulants} \\

\vspace{4mm}

I.P.~Lokhtin$^\dag$,  
L.I.~Sarycheva$^\ddag$ and 
A.M.~Snigirev$^\S$  \\
M.V.Lomonosov Moscow State University, D.V.Skobeltsyn Institute of Nuclear Physics \\
119992, Leninskie Gory, Moscow, Russia 
\end{center}  

\begin{abstract}  
We analyze the method for calculation of a coefficient of jet azimuthal 
anisotropy without reconstruction of the nuclear reaction plane considering 
the higher order correlators between the azimuthal position of jet axis and 
the angles of particles not incorporated in the jet. The reliability of this 
technique in the real physical situation under LHC conditions is illustrated.  
\end{abstract}

\bigskip

\vspace{100mm}
\noindent
---------------------------------------------\\
$^\dag$ E-mail: igor@lav01.sinp.msu.ru\\
$^\ddag$ E-mail: lis@alex.sinp.msu.ru\\ 
$^\S$ E-mail: snigirev@lav01.sinp.msu.ru\\
\end{titlepage}   
\newpage 

\section {Introduction}   

Nowadays the strong interest is springing up to the investigations and measurements 
of azimuthal correlations in ultrarelativistic heavy ion collisions 
(see, for instance,~\cite{qm01} and references therein). One of the main 
reasons is that the rescattering and energy loss of hard partons in the azimuthally 
non-symmetric volume of dense quark-gluon matter, created initially in the nuclear 
overlap zone in collisions with non-zero impact parameter, can result in the visible 
azimuthal anisotropy of high-$p_T$ hadrons at RHIC~\cite{qm01,uzhi,wang00,gyul00}.

Recent anisotropic flow data at RHIC~\cite{star,phenix,phobos} can be described well by 
hydrodynamical models for semi-central collisions and $p_T$ up to $\sim 2$ 
GeV/c (the elliptic flow coefficient $v_2$ appears to be monotonously growing with 
increasing $p_T$~\cite{kolb} in this case), while the majority of microscopical Monte-Carlo models 
underestimate the flow effects (see however~\cite{zabrodin}). The saturation and 
gradual decrease of $v_2$ at relatively large transverse momentum ($p_T \ga 2$ GeV/c), 
predicted as a signature of strong partonic energy loss in a dense QCD plasma, seem now 
to be supported by the preliminary 
data extending up to $p_T \simeq 10$ GeV/c at RHIC. The interpolation between the 
low-$p_T$ relativistic hydrodynamics region and the high-$p_T$ pQCD-computable region 
was evaluated in~\cite{gyul00}. 

The initial gluon densities in Pb$-$Pb reactions at $\sqrt{s}_{NN}=5.5$~TeV   
at the Large Hadron Collider (LHC) are expected to be significantly higher
than at RHIC, implying stronger partonic energy loss. Moreover, since at LHC
energies the  inclusive cross section for hard jet production at   
$E_T \sim 100$ GeV is large enough to study the impact parameter 
dependence of such processes~\cite{lokhtin00}, one can hope to observe the
azimuthal anisotropy for hadronic jet itself~\cite{lokhtin01,Lokhtin:2002}.  
In particular, CMS 
experiment at LHC~\cite{cms94} will be able to provide both the jet reconstruction 
and adequate measurement of impact parameter of nuclear collision using 
calorimetric information~\cite{note00-060}. In the case of jets, the methodical 
advantage of azimuthal observables is that one 
needs to reconstruct only azimuthal position of the jet without measuring its total  
energy. It can be done more easily and with high accuracy, while the 
reconstruction of jet energy is a more ambiguous problem~\cite{note00-060}.
However the measurement of jet production as a function of azimuthal angle 
requires event-by-event determination of the nuclear event plane based on
the anisotropic flow analysis.  

The methods for elliptic flow analysis can be generally divided in two categories:
two-particle methods suggested and summarized in works~\cite{voloshin,wang,ollit} and 
multi-particle methods~\cite{ollitr,din}. In two-particle methods the contribution of   
non-flow (non-geometric) correlations to the determination of the elliptic flow
coefficient $v_2$ is of the order of $1/\sqrt{N_0}$, where $N_0$ is the measured 
multiplicity. In multi-particle methods this contribution goes down typically as 
$1/N_0^{3/4}$, i.e., smaller by a factor of the order of $N_0^{1/4}$. Thus experimental 
techniques based on higher order cumulant analysis should be able in many cases to 
allow access to the smaller values of azimuthal particle anisotropy in comparison with 
two-particle methods, due to automatic elimination of the major non-flow many-particle 
correlations and the systematic errors originating from azimuthal asymmetry of the 
detector acceptance.
This kind of analysis for particle flow has been already done by the 
STAR Collaboration at RHIC~\cite{star02}.

In our previous Letter~\cite{phl02} we proposed the method for measurement of 
jet azimuthal anisotropy coefficients without direct reconstruction of the 
event plane, and illustrated its reliability in a real experimental situation.
This technique is based on the calculation of correlations between the azimuthal 
position of the jet axis and the angles of particles not incorporated in the jet, the 
azimuthal distribution of jets being described by the elliptic form. To improve our 
approach, in the given paper we extend our analysis~\cite{phl02}, considering the 
cumulant expansion~\cite{ollitr} of multi-particle azimuthal correlations. 

\section{Correlators versus the jet elliptic anisotropy}

Let us remind some features of our previous investigation~\cite{phl02}.
We start from the essence of techniques~\cite{voloshin, wang, ollit} for 
measuring azimuthal elliptic anisotropy of particle distribution, which can be 
written in the form  
\begin{equation} 
\label{phi_part}
\frac{dN}{d \varphi} = \frac{N_0}{2\pi}~ [1+2v_2\cos{2 
(\varphi -\psi_R)}]~,~~~~
 N_0 = \int\limits_{-\pi}^{\pi}d\varphi~\frac{dN}{d \varphi}~ . 
\end{equation}  
Knowing the nuclear reaction plane angle $\psi_R$ allows one to determine the 
coefficient $v_2$ of azimuthal anisotropy of particle flow  as an average (over
particles) cosine of $2 \varphi$: 
\begin{equation} 
\label{v2_part}
 < \cos{2(\varphi-\psi_R)} >~=~
 \frac{1}{N_0}~\int\limits_{-\pi}^{\pi}d\varphi~\cos{2(\varphi -\psi_R)}~
 \frac{dN}{d \varphi}~=~v_2~.
\end{equation}      
However in the case when there are no other correlations of particles except 
those due to flow (or such other correlations can be neglected), the 
coefficient of azimuthal anisotropy can be determined using two-particle
azimuthal correlator without the event plane angle $\psi_R$: 
\begin{eqnarray} 
\label{v2^2_part}
& & < \cos{2(\varphi_1-\varphi_2)} >~=~
\frac{1}{N_0^2}~\int\limits_{-\pi}^{\pi}d\varphi_1
~\int\limits_{-\pi}^{\pi}d\varphi_2
 ~\cos{2(\varphi_1 -\varphi_2)}~
 \frac{d^2N}{d \varphi_1d\varphi_2}~\nonumber \\
& & =\frac{1}{N_0^2}~\int\limits_{-\pi}^{\pi}d\varphi_1
~\int\limits_{-\pi}^{\pi}d\varphi_2
 ~\cos{2((\varphi_1 -\psi_R) -(\varphi_2 -\psi_R))}~
 \frac{dN}{d \varphi_1}~\frac{dN}{d\varphi_2}~=~v_2^2~.
\end{eqnarray}   

Let us consider now the event with high-p$_T$ jet (dijet) production, the 
distribution of jets over azimuthal angle relatively to the reaction plane 
being described well by the elliptic form~\cite{lokhtin01},  
\begin{equation} 
\label{phi_jet}
\frac{dN^{jet}}{d \varphi} = \frac{N^{jet}_0}{2\pi}~ [1+2v^{jet}_2\cos{2 
(\varphi -\psi_R)}]~,~~~~
 N^{jet}_0 = \int\limits_{-\pi}^{\pi}d\varphi~\frac{dN^{jet}}{d \varphi}~, 
\end{equation}
where the coefficient of jet azimuthal anisotropy $v^{jet}_2$ is determined  
as an average over all events cosine of $2 \varphi$, 
\begin{equation} 
\label{u2_jet}
 \left< \cos{2(\varphi-\psi_R)} \right>_{event}~=~
 \frac{1}{N^{jet}_0}~\int\limits_{-\pi}^{\pi}d\varphi~\cos{2(\varphi -\psi_R)}~
 \frac{dN^{jet}}{d \varphi}~=~v^{jet}_2~. 
\end{equation} 
One can calculate the correlator between the azimuthal position of jet axis 
$\varphi _{jet}$\footnote{The other possibility is to fix the azimuthal position of 
a leading particle in the jet. In this case calculating azimuthal correlations can 
provide the information about azimuthal anisotropy of high-$p_T$ particle spectrum.} 
and the angles of particles, which are not incorporated in the jet(s). The value of 
this correlator is related to the elliptic coefficients $v_2$ and $v^{jet}_2$ as  
\begin{eqnarray} 
\label{cor}
\left< < \cos{2(\varphi _{jet}-\varphi)} >\right>_{\rm event} & = & 
 \frac{1}{N^{jet}_0N_0}~\int\limits_{-\pi}^{\pi}d\varphi _{jet}~
\int\limits_{-\pi}^{\pi} d\varphi~
\cos{2(\varphi _{jet} -\varphi)}~ \frac{dN^{jet}}{d
\varphi_{jet}}~\frac{dN}{d\varphi}\nonumber \\ 
& = & \frac{1}{N^{jet}_0}~\int\limits_{-\pi}^{\pi}d\varphi _{jet}~
\cos{2(\varphi _{jet} -\psi _R)}~ \frac{dN^{jet}}{d
\varphi_{jet}}~v_2~=~v^{jet}_2~v_2~. 
\end{eqnarray}
 
Using Eq. (\ref{v2^2_part}) and intermediate result in Eq. (\ref{cor}) (after averaging
over particles $\cos{2(\varphi _{jet} -\varphi)}$ reduces to $v_2 \cos{2(
\varphi _{jet} -\psi _R)}$) we derive the formula for computing absolute value 
of the coefficient of jet azimuthal anisotropy (without reconstruction of sign of 
$v^{jet}_2$): 
\begin{equation} 
\label{u2_event}
 v^{jet}_2 ~=~ \left< \frac{< \cos{2(\varphi _{jet}-\varphi)} >}
 {\sqrt{< \cos{2(\varphi_1-\varphi_2)} >}}
 \right>_{\rm event}~.
\end{equation}
This formula does not require the direct determination of reaction plane angle 
$\psi_R$. The brackets $\left< ~~~~\right>$ represent the averaging over
particles (not incorporated in the jet) in a given event, while the brackets 
$\left< ~~~~\right>_{event}$ are the averaging over events. 

The formula (\ref{u2_event}) can be generalized by introducing as weights the
particle momenta, 
\begin{equation} 
\label{u2(p)_event}
 v^{jet}_{2(p)} ~=~ \left< \frac{<~p_T(\varphi) \cos{2(\varphi _{jet}-\varphi)} 
>} {\sqrt{<~p_{T1}(\varphi_1)~p_{T2}(\varphi_2)~ \cos{2(\varphi_1-\varphi_2)}
 >}}
 \right>_{\rm event}~.
\end{equation}
In this case the  brackets $\left< ~~~~\right>$ denote the averaging over angles
and transverse momenta of particles. The other modification of  
(\ref{u2(p)_event}), 
\begin{equation} 
\label{u2(E)_event}
 v^{jet}_{2(E)} ~=~ \left< \frac{<~E(\varphi) \cos{2(\varphi _{jet}-\varphi)}>}
 {\sqrt{<~E_1(\varphi_1)~E_2(\varphi_2) \cos{2(\varphi_1-\varphi_2)}>}}
 \right>_{\rm event}~,
\end{equation} 
($E_i (\varphi_i)$~being energy deposit in a calorimetric segment $i$ 
of position $\varphi_i$) allows one using calorimetric measurements 
(\ref{u2(E)_event}) for the determination of jet azimuthal anisotropy. 

\section{Higher order correlators}

The main advantage of the higher order cumulant analysis is in the fact that, 
as argued in Ref.~\cite{ollitr}, if flow is larger than non-flow correlations, 
the contribution of the latter to $v_2$ extracted from higher order correlators 
is suppressed\footnote{This can be essential under data analysis 
with not high enough multiplicity of particles in an event.} by powers of 
particle multiplicity $N_0$ in an event. 

Thus, for example, the fourth order cumulant for elliptic particle flow is 
defined as~\cite{ollitr}
\begin{eqnarray} 
\label{cum}
c_2[4]~~\equiv~~< \cos{2(\varphi_1+\varphi_2 -\varphi_3-\varphi_4)} >~~~~~~~~~~~~~~~~
~~~~~~~~~~~~~~~~\nonumber\\
-~< \cos{2(\varphi_1 -\varphi_3)} >~< \cos{2(\varphi_2 
-\varphi_4)} >
-~< \cos{2(\varphi_1 -\varphi_4)} >~< \cos{2(\varphi_2 
-\varphi_3)} > ~,
\end{eqnarray}  
and in the case of existing only correlations with the reaction plane (i.e. 
the factorization of multi-particle distributions is held as in Eq.~(\ref{v2^2_part})) 
is equal to
\begin{equation} 
\label{cum-v_2}
  c_2[4]~=~-v_2^4.
\end{equation} 

If now one defines the coefficient $v_2$ of azimuthal anisotropy through 
two-particle correlator
\begin{equation} 
\label{v_2-cor}
  v_2~=~ \sqrt{< \cos{2(\varphi_1-\varphi_2)}>},
\end{equation} 
then the contribution of non-flow correlations, as argued in~\cite{ollitr},
is of order  $1/\sqrt{N_0}$. While their contribution to $v_2$, extracted from 
the fourth order correlator 
\begin{equation} 
\label{v_2-cor4}
  v_2~=~ (-c_2[4])^{1/4},
\end{equation} 
scales as $1/N_0^{3/4}$, i.e. it is suppressed by an extra factor of $1/N_0^{1/4}$.
Corresponding data analysis based on Eq.~(\ref{cum}) and result (\ref{cum-v_2}) 
has been already carried out at STAR~\cite{star02}.
  
Now using the derivation of Eq.~(\ref{u2_event}) and the result 
(\ref{cum-v_2}) it is straightforward to obtain the formula for calculation
(measurement) of coefficient of jet azimuthal anisotropy through the higher 
order correlator, which is less sensitive to non-flow correlations: 
\begin{eqnarray} 
\label{v2(4)}
v_2^{jet}[4]~=\bigg\langle\frac{1}{(-c_2[4])^{3/4}}
[~-< \cos{2(\varphi_{jet}+\varphi_1 -\varphi_2
-\varphi_3)} >~\nonumber\\
~~~~~~~~~+~< \cos{2(\varphi_{jet} -\varphi_2)} >~
~< \cos{2(\varphi_1 
-\varphi_3)} >~\nonumber\\
~~~~~~~~~+~< \cos{2(\varphi_{jet} -\varphi_3)} >~~< \cos{2(\varphi_1 
-\varphi_2)} >]\bigg\rangle_{\rm event}.
\end{eqnarray}  
Here we stress once more that in the case of existing only correlations with 
the reaction plane, Eq.~(\ref{v2(4)}) together with Eq.~(\ref{u2_event}) 
transforms into identity. The formula just derived can be, as in Sect.~2, 
generalized for calorimetric measurements of energy flows:
\begin{eqnarray} 
\label{v2E(4)}
& &v_{2(E)}^{jet}[4]~=\bigg\langle\frac{1}{(-c_{2(E)}[4])^{3/4}}[~-< E_{1}(\varphi_1)
E_{2}(\varphi_2)E_{3}(\varphi_3)
\cos{2(\varphi_{jet}+\varphi_1 -\varphi_2
-\varphi_3)} >\nonumber\\
& &~~~~~~~~~+~< E_{2}(\varphi_2)\cos{2(\varphi_{jet} -\varphi_2)} >~
~< E_{1}(\varphi_1)E_{3}(\varphi_3)\cos{2(\varphi_1 
-\varphi_3)} >\nonumber\\
& &~~~~~~~~~+~< E_{3}(\varphi_3)\cos{2(\varphi_{jet} 
-\varphi_3)} >~~<E_{1}(\varphi_1) E_{2}(\varphi_2)\cos{2(\varphi_1 
-\varphi_2)} >]~\bigg\rangle_{\rm event},
\end{eqnarray}  
where
\begin{eqnarray} 
\label{cumE}
& & c_{2(E)}[4]~=~<E_{1}(\varphi_1)E_{2}(\varphi_2)E_{3}(\varphi_3)
E_{4}(\varphi_4) 
\cos{2(\varphi_1+\varphi_2 -\varphi_3-\varphi_4)} >\nonumber\\
& &-~<E_{1}(\varphi_1) E_{3}(\varphi_3)\cos{2(\varphi_1 -\varphi_3)} >~~
<E_{2}(\varphi_2)E_{4}(\varphi_4) \cos{2(\varphi_2 -\varphi_4)} >\nonumber\\
& &-~< E_{1}(\varphi_1)E_{4}(\varphi_4)\cos{2(\varphi_1 -\varphi_4)} >~~
<E_{2}(\varphi_2)E_{3}(\varphi_3) \cos{2(\varphi_2 -\varphi_3)} >.
\end{eqnarray} 

In the case when the azimuthal position of jet axis correlates not only with 
the reaction plane, one can try to improve this technique using the multiple
correlators of another form: averaging over not all events but selecting some
their sub-events. For instance, one can consider sub-events $1$ and $2$, when jets 
are produced with the rapidity $y>0$ and $y<0$. Then calculating correlator  
\begin{eqnarray} 
\label{c2jet(4)}
c_2^{jet}[4]~=\bigg\langle\frac{1}{\sqrt{<\cos{2(\varphi_1-\varphi_2)}>
<\cos{2(\phi_1-\phi_2)}>}}
[~< \cos{2(\varphi_{jet}-\varphi +\phi_{jet}-\phi)} >\nonumber\\
+< \cos{2(\varphi_{jet}-\varphi -\phi_{jet}+\phi)} >\nonumber\\
~~~~~~~~~-~< \cos{2(\varphi_{jet} -\varphi)} >~~< \cos{2(\phi_{jet} 
-\phi)} >]\bigg\rangle_{{\rm sub-event}~1,~2},
\end{eqnarray}  
we obtain that, if there are flow particle correlations only and 
the distribution of jets over azimuthal angles is described by the 
elliptic form (\ref{phi_jet}) in every sub-event, it is equal to 
\begin{equation} 
\label{cum(jet)-v_2}
  c_2^{jet}[4]~=~v_2^{jet}(y>0)~v_2^{jet}(y<0).
\end{equation}
In Eq.~(\ref{c2jet(4)}) the angles $\varphi$ are defined as the azimuthal angles 
of particles and jets in sub-event with $y>0$, and  $\phi$ --- in sub-event with
$y<0$. 
Correspondingly the brackets $<~~~~>$ represent the averaging over particles in
sub-events 1, 2, while the brackets
 $\bigg\langle~~~\bigg\rangle_{\rm sub-event~1,~2}$ are the averaging over these
 sub-events. The generalization of Eq.
(\ref{c2jet(4)}) for calorimetric measurements of energy flow is obvious
(similar to Eqs.~(\ref{u2(E)_event}) and (\ref{v2E(4)})). We do not write also 
this result specially as the examples of utilizing six- and other higher order
correlators.

\newpage

\section{Non-flow correlations}

Here we discuss the influence of non-flow correlations\footnote{See also
Appendix of work~\cite{ollitr} and Ref.~\cite{kov}} on the $v_2^{jet}$
determination. There are various sources of such correlations, among which
minijet production~\cite{kov}, global momentum conservation~\cite{dan, bor02}, 
resonance decays (in which the decay products are correlated), final state 
Coulomb, strong or quantum interactions~\cite{bor}. We restrict our 
consideration to two-particle correlations only. It will be enough to 
illustrate the advantage in using higher order cumulants. In this case 
multi-particle distributions are not factorized again and instead 
of Eq.~(\ref{v2^2_part}) we have
\begin{eqnarray} 
\label{c2^2_part}
& & c_2[2]~\equiv~< \cos{2(\varphi_1-\varphi_2)} >~=~
\frac{1}{N_2}~\int\limits_{-\pi}^{\pi}d\varphi_1
~\int\limits_{-\pi}^{\pi}d\varphi_2
 ~\cos{2(\varphi_1 -\varphi_2)}~
 [\frac{dN}{d \varphi_1}~\frac{dN}{d \varphi_2}~+~
  \frac{dN_{cor}}{d \varphi_1d\varphi_2}]\nonumber \\
& &~~~~~~~~~~~~~~~~~~~ =~v_2^2~\frac{1~+~v_{cor}^{-}/v_2^2}{1~+~\Delta}~,
\end{eqnarray}   
where
\begin{eqnarray} 
\label{norma}
& & N_2~=~
~\int\limits_{-\pi}^{\pi}d\varphi_1
~\int\limits_{-\pi}^{\pi}d\varphi_2~
 [\frac{dN}{d \varphi_1}~\frac{dN}{d \varphi_2}~+~
  \frac{dN_{cor}}{d \varphi_1d\varphi_2}]~,\nonumber \\
& & \Delta~=~\frac{1}{N_0^2}
~\int\limits_{-\pi}^{\pi}d\varphi_1
~\int\limits_{-\pi}^{\pi}d\varphi_2
~~  \frac{dN_{cor}}{d \varphi_1d\varphi_2}~,\nonumber \\
& & v_{cor}^{-}~=~\frac{1}{N_0^2}
~\int\limits_{-\pi}^{\pi}d\varphi_1
~\int\limits_{-\pi}^{\pi}d\varphi_2
 ~\cos{2(\varphi_1 -\varphi_2)}~
 ~~ \frac{dN_{cor}}{d \varphi_1d\varphi_2}~.
\end{eqnarray}
Eq.~(\ref{cor}) remains unchangeable and result (\ref{u2_event}) transforms 
into
\begin{equation} 
\label{v2_event}
 v^{jet}_2[2] ~\equiv~ \left< \frac{< \cos{2(\varphi _{jet}-\varphi)} >}
 {\sqrt{< \cos{2(\varphi_1-\varphi_2)} >}}
 \right>_{event}~=~v_2^{jet}~
\sqrt{ \frac{1~+~\Delta}{1~+~v_{cor}^{-}/v_2^2}}~.
\end{equation}

After some algebra the fourth order cumulant (\ref{cum}) reduces to
\begin{eqnarray} 
\label{c_2[4]}
& &  c_2[4] ~=~
~v_2^4~\frac{1~+~4v_{cor}^{-}/v_2^2~+~
2v_{cor}^{-~-}/v_2^4~+~2v_{cor}^{+}/v_2^2~
+~v_{cor}^{+~+}/v_2^4 }{1~+~6\Delta~+~3\Delta^2}\nonumber \\
& &  
-2v_2^4~\left(\frac{1~+~v_{cor}^{-}/v_2^2}{1~+~\Delta}\right)^2,
\end{eqnarray}
where
\begin{eqnarray} 
\label{v_cor+}
& & v_{cor}^{+}~=~\frac{1}{N_0^2}
~\int\limits_{-\pi}^{\pi}d\varphi_1
~\int\limits_{-\pi}^{\pi}d\varphi_2
~\cos{2(\varphi_1-\psi_R +\varphi_2-\psi_R)}~
 ~ \frac{dN_{cor}}{d \varphi_1d\varphi_2}~,\nonumber\\ 
& & v_{cor}^{+~+}~=~\frac{1}{N_0^4}
~\int\limits_{-\pi}^{\pi}d\varphi_1
~\int\limits_{-\pi}^{\pi}d\varphi_2
~\int\limits_{-\pi}^{\pi}d\varphi_3
~\int\limits_{-\pi}^{\pi}d\varphi_4
 ~\cos{2(\varphi_1 +\varphi_2-\varphi_3 - \varphi_4)}~
~ \frac{dN_{cor}}{d \varphi_1d\varphi_2}~\frac{dN_{cor}}{d \varphi_3d\varphi_4}, 
\nonumber\\
& & v_{cor}^{-~-}~=~\frac{1}{N_0^4}
~\int\limits_{-\pi}^{\pi}d\varphi_1
~\int\limits_{-\pi}^{\pi}d\varphi_2
~\int\limits_{-\pi}^{\pi}d\varphi_3
~\int\limits_{-\pi}^{\pi}d\varphi_4
 ~\cos{2(\varphi_1 +\varphi_2-\varphi_3 - \varphi_4)}~
~ \frac{dN_{cor}}{d \varphi_1d\varphi_3}~\frac{dN_{cor}}{d \varphi_2d\varphi_4}
\nonumber\\
& & =~(v_{cor}^{-})^2~-~\left(\frac{1}{N_0^2}
~\int\limits_{-\pi}^{\pi}d\varphi_1
~\int\limits_{-\pi}^{\pi}d\varphi_3
~\sin{2(\varphi_1-\varphi_3)}~
 ~ \frac{dN_{cor}}{d \varphi_1d\varphi_3}\right)^2~
\end{eqnarray}

$=~(v_{cor}^{-})^2$~ if ~$\frac{dN_{cor}}{d \varphi_1d\varphi_2}$ ~is an even
function of angular difference $(\varphi_1-\varphi_2)$.

\noindent
The numerator in Eq.~(\ref{v2(4)}) is rewritten in the following form:
\begin{eqnarray} 
\label{Num}
& & Num~\equiv~
-< \cos{2(\varphi_{jet}+\varphi_1 -\varphi_2
-\varphi_3)} >~\nonumber \\
& & +< \cos{2(\varphi_{jet} -\varphi_2)} >< \cos{2(\varphi_1 
-\varphi_3)} >
+< \cos{2(\varphi_{jet} -\varphi_3)} >< \cos{2(\varphi_1 
-\varphi_2)} >\nonumber \\
& & =
~-~v_2^3\cos{2(\varphi_{jet}-\psi_R)}~\frac{1~+~2v_{cor}^{-}/v_2^2~+~
v_{cor}^{+}/v_2^2~ }{1~+~3\Delta}\nonumber \\
& &  
+2v_2^3~\cos{2(\varphi_{jet}-\psi_R)}~\frac{1~+~v_{cor}^{-}/v_2^2}{1~+~\Delta}~
+~SIN,
\end{eqnarray}  
where the terms ~$SIN$~ are proportional to ~$\sin{2(\varphi_{jet} -\psi_R)}$~ 
and vanishing after averaging over $\varphi_{jet}$.

At the first glance it is hard to see the advantage in using higher order 
cumulants from Eqs.~(\ref{v2_event}), (\ref{c_2[4]}) and (\ref{Num}). However, it 
is reasonable to suppose that the contribution of two-particle correlations to 
the normalization factor $N_2$ is small, $\Delta \ll 1$, while their ``second 
Fourier harmonic'' $v_{cor}^{-}$ can be of the order of $v_2^2$. Then {\it all}
direct two-particle correlations $v_{cor}^{-}$ are automatically {\it canceled
out}\footnote{This is one of the main advantage of the cumulant expansion.} from
Eqs.~(\ref{c_2[4]}) and (\ref{Num}) in first order in $\Delta$,
but are {\it survived} in Eq.~(\ref{v2_event}). The non-direct two-particle 
correlations $v_{cor}^{+},~v_{cor}^{+~+}$ are survived. But they are suppressed in
the comparison with the direct correlations $v_{cor}^{-}$ (contributing to
$v_2^{jet}[2]$) due to the fact that $\frac{dN_{cor}}{d \varphi_1d\varphi_2}$
is an even function of angular difference
$(\varphi_1 - \varphi_2)$ only in most physically interesting
cases~\cite{kov}. Moreover, for the small-angle $\delta$-like correlations
($\frac{dN_{cor}}{d \varphi_1d\varphi_2}~\sim~\frac{\exp(-(\varphi_1-
\varphi_2)^2/2\sigma^2)}{\sqrt{2\pi}\sigma},~ \sigma \rightarrow 0$) and for
the large-angle oscillating ones 
($\frac{dN_{cor}}{d \varphi_1d\varphi_2}~\sim~\cos2(\varphi_1-
\varphi_2)$) the non-direct correlations $v_{cor}^{+},~v_{cor}^{+~+}$ are 
equal to zero. Then 
\begin{eqnarray} 
\label{v2(4)`}
v_2^{jet}[4]~=\bigg\langle\frac{Num}{(-c_2[4])^{3/4}}
\bigg\rangle_{event}~\simeq~ v_2^{jet}
\end{eqnarray}  
in this case, while 
\begin{equation} 
\label{v2_event`}
 v^{jet}_2[2] ~\simeq~v_2^{jet}~
\sqrt{ \frac{1}{1~+~v_{cor}^{-}/v_2^2}}~.
\end{equation}
Thus Eqs.~(\ref{v2(4)`}) and (\ref{v2_event`}) demonstrate the better accuracy of 
higher order cumulants explicitly.
 
\section{Discussion}
In order to illustrate the applicability of the method presented with regard for the 
real physical situation, we consider the following model (see Ref.~\cite{phl02} for 
details).  

\smallskip 

\centerline{\bf The model} 

The initial jet distributions in a nucleon-nucleon sub-collision at $\sqrt{s}=5.5$ TeV 
have been generated using PYTHIA$\_5.7$~\cite{pythia}. We simulated the  
rescattering and energy loss of jets in gluon-dominated plasma, created initially in 
the nuclear overlap zone in Pb$-$Pb collisions at different impact parameters. 
For details of this approach one can refer to Refs.~\cite{lokhtin00,lokhtin01}. 
Essential for our consideration here is that   
in non-central collisions the azimuthal distribution of jets is approximated well
by the elliptic form (\ref{phi_jet}). In the model the coefficient of jet azimuthal 
anisotropy increases almost linearly with the growth of impact parameter $b$ and 
becomes maximum at $b \sim 1.2 R_A$, where $R_A$ is the nucleus radius. After that 
$v_2^{jet}$ drops rapidly with increasing $b$: this is the domain of impact parameter 
values, where the effect of decreasing energy loss due to reducing effective transverse 
size of the dense zone and initial energy density of the medium is crucial and not 
compensated anymore by stronger non-symmetry of the volume. 
The kinematical cuts on jet transverse energy and rapidity has been applied: 
$E_T^{jet} > 100$ GeV and  $|y^{jet}| < 1.5$.  After this the dijet event is 
superimposed on the Pb$-$Pb event containing anisotropic flow.  

Anisotropic flow was generated using the simple hydrodynamical Monte-Carlo 
code~\cite{kruglov,phl02} giving hadron (charged and neutral pion, kaon 
and proton) spectrum as a superposition of the thermal distribution and collective 
flow. To be definite, we fixed the following ``freeze-out'' parameters: the 
temperature $T_f = 140$ MeV, the collective  longitudinal rapidity $Y_L^{max}=3$ and 
the collective transverse rapidity $Y_T^{max}=1$. We set the Poisson multiplicity 
distribution and took into account the impact parameter dependence of multiplicity in a 
simple way, just suggesting that the mean multiplicity of particles is proportional to 
the nuclear overlap function. We also suggested~\cite{phl02} that the spatial 
ellipticity of the ``freeze-out'' region is directly related to the initial spatial 
ellipticity of the nuclear overlap zone. 
Such ``scaling'' allows one to avoid using additional parameters
and, at the same time, results in the elliptic anisotropy of particle and
energy flow due to the dependence of effective transverse size of the ``freeze-out''
region on the azimuthal angle of a ``hadronic liquid'' element. 
Obtained in such a way azimuthal distribution of particles is described well by 
the elliptic form (\ref{phi_part}) for the domain of reasonable impact parameter
values.  

To be specific, we consider the geometry of CMS detector~\cite{cms94} at LHC. 
The central (``barrel'') part of the CMS calorimetric 
system  covers the pseudo-rapidity region $|\eta| < 1.5$, the segmentation 
of electromagnetic and hadron calorimeters being $\Delta \eta \times \Delta 
\phi = 0.0174 \times 0.0174$ and $\Delta \eta \times \Delta \phi = 0.0872 \times 
0.0872$ respectively~\cite{cms94}. In order to reproduce roughly the experimental 
conditions (not including real detector effects, but just assuming 
calorimeter hermeticity), we applied Eqs.~(\ref{u2(E)_event}) and (\ref{v2E(4)}) to the 
energy deposition $E_i(\varphi _i)$ of generated particles, integrated over the 
rapidity in $72$ segments (according to the number of segments in the hadron 
calorimeter: $72~\times~0.0872~=~2\pi$; $i~=~1,...,72$) covering full azimuth. 

Note that in the CMS heavy ion physics program, the modified sliding window-type jet 
finding algorithm has been developed to search for ``jet-like'' clusters 
above the average energy, and to subtract the background from the underlying 
event~\cite{note00-060}. Strictly speaking, after jet extraction the background energy 
deposition in the calorimetric cells should be redefined and can appear to be not 
exactly equal to the initially generated one. However we neglect this effect here. In a 
real experimental situation, in order to avoid the influence of jet contribution on the 
particle flow, one can consider jets and particles incorporated in the energy flow 
analysis in different rapidity regions.                                      

\smallskip 

\centerline{\bf Numerical results}

We have found~\cite{phl02} that the accuracy of $v^{jet}_2$
determination from Eq.~(\ref{u2(E)_event}) is close to $100 \%$ for semi-central 
($b~\la~R_A$) collision and becomes significantly 
worse in very peripheral collision ($b \sim 2 R_A$), wherein decreasing 
multiplicity and azimuthal anisotropy of the event results in  relatively  large
fluctuations of energy deposition in each segment.

In the given paper we test the efficiency of the higher order correlator 
(\ref{v2E(4)}) and have found, at first, that the results for $v_2^{jet}$ 
obtained from Eqs.~(\ref{u2(E)_event}) and (\ref{v2E(4)}) are practically the 
same. This is explained by the fact that our simple Monte-Carlo event generator gives 
the elliptic anisotropy of energy flow, correlated with the reaction plane, but 
no correlations between energy deposition in the calorimeter 
segments. We can introduce such correlations at the calorimetric level ``by 
hand'', simply assuming that the probability of finding the energy $E_i$ in a 
segment $i$ and the energy $E_j$ in a segment $j$  is proportional to 
$E_i E_j(1~+~c_{ij})$, where the ``correlation strength'' $c_{ij}$ may be, for example,  
proportional to $\delta_{ij}$ (the short-range $\delta$-like correlations) or 
$\cos{2(\varphi_i-\varphi_j)}$ (the long-range oscillating correlations). In 
this case we became convinced  that the higher order cumulant (\ref{v2E(4)}) 
was almost independent of such correlations (as it was shown in Sect. 4), 
while the result of calculation (\ref{u2(E)_event}) changed, closely 
following the formula (\ref{v2_event`}) corrected by autocorrelation terms which
are non-vanishing in finite summation~\cite{ollitr}. 

We have also found at the calorimetric level that, taking into account the 
effect of possible detector inefficiency (i.e. that the particles and jets are not
detected  in a ``blind'' azimuthal sector of size $\alpha$), 
the accuracy of $v_2^{jet}$ determination appears to be less than 
$50 \%$ at $\alpha~\ga 30^{\circ}$ in our model calculation without correlations 
and at $b \ge R_A$, whichever 
algorithm (\ref{u2(E)_event}) or (\ref{v2E(4)}) we used.

Fig.1 is presented to illustrate the improvement due to the fourth order cumulant 
method in the determination of jet azimuthal anisotropy $v^{jet}_2$ depending on
the ratio $\bar v^-_{cor}/\bar v_2^2$, where $\bar v_2 $ is the coefficient of
elliptic azimuthal anisotropy of energy flow defined here as 
\begin{equation} 
 \bar v_2 ~=~\frac{1}{2}~~ \frac{E_{max(i)} - E_{min(i)}}
 {E_{max(i)} + E_{min(i)}}~, 
\end{equation} 
and $E_{max(i)}$ and $E_{min(i)}$ are the maximum and minimum energy deposit in a 
segment respectively ($i~=~1,...,72$). The coefficient $\bar v^-_{cor}$ determines the 
``correlation strength'' at the calorimetric level, $c_{ij} = 72 \bar v^-_{cor} 
\delta_{ij}$ for short-range correlations 
\footnote{
Here one should note that the majority of sources of non-flow correlations mentioned 
above is effective at small angles between particles. In our case these correlations 
can be partially smoothed out after summing particle energies over the azimuthal angles 
in a calorimeter segment of finite size ($\sim 5^{\circ}$). This can result in the
smaller value of the ratio $\bar v^-_{cor}/\bar v_2^2$ in comparison with the ratio 
$v^-_{cor}/v_2^2$ (and, as consequence, in a somewhat less improvement due to the higher
order method at the calorimetric level (\ref{v2E(4)}) in comparison with the
particle level (\ref{v2(4)})). We still have no any adequate Monte-Carlo generator 
for particle flow effects at LHC including the physical model for correlations. Thus   
we can not estimate the real value of $\bar v^-_{cor}/\bar v_2^2$ (true benefit  
from the higher order method) and just treat it here as a phenomenological parameter.}
(the similar result is obtained for long-range correlations with 
$c_{ij} = 2 \bar v^-_{cor} \cos{2(\varphi_i-\varphi_j)}$). 
The plots show the $b$-dependence of the ``theoretical'' value of $v^{jet}_2$ 
(calculated including collisional and radiative energy loss
when the reaction plane angle is known in each event), and the $v^{jet}_2$ 
determined by the methods (\ref{u2(E)_event}) and (\ref{v2E(4)}) 
for the three ratios $\bar v^-_{cor}/ \bar v_2^2 ~=~ 0,~ 0.01,~0.1$. We 
used two values of the input parameter, the number of charged particles per unit 
rapidity at $y=0$ in central  Pb$-$Pb collisions: $dN^{\pm}/dy = 3000$ (Fig.1a) and 
$6000$ (Fig.1b).  

One can see that improvement due to the fourth order cumulant method (result of 
(\ref{v2E(4)}) is independent of $\bar v^-_{cor}/\bar v_2^2$ and   
coincides with the result of (\ref{u2(E)_event}) for $\bar v^-_{cor}/\bar v_2^2=0$) is 
pronounced for more peripheral collisions, smaller particle multiplicities and
larger ``correlation strengths''.   

\newpage 

\section{Conclusions}

In the present paper we have analyzed the method for measurements of 
jet azimuthal anisotropy coefficients without reconstruction of the event 
plane considering the higher order correlators between the azimuthal position 
of the jet axis and the angles of particles not incorporated in the jet.
The method is generalized by introducing as weights the particle momenta or 
the energy deposit in the calorimeter segments. In the latter case, we have 
illustrated its reliability in the real physical situation under LHC conditions. 
Introducing in the model correlations between energy deposits in the 
calorimeter segments does not practically change the accuracy of the method using 
fourth order cumulant calculations (\ref{v2E(4)}), while the result obtained 
with second order correlator (\ref{u2(E)_event}) is dependent significantly of  
the ``strength'' of such correlations. The advantage of the higher order cumulant
analysis is pronounced for more peripheral collisions and smaller particle 
multiplicities.

To summarize, we believe that the present technique may be useful 
investigating azimuthal anisotropy of jets and high-$p_T$
particles in heavy ion collisions at RHIC and LHC. 

{\it Acknowledgments}.  
Discussions with A.I.~Demianov, S.V.~Petrushanko, V.V.~Uzhinskii, I.N.~Vardanian, 
S.A.~Voloshin, U.~Wiedemann and E.E.~Zabrodin are 
gratefully acknowledged.

\newpage 

\begin{figure}[hbtp] 
\begin{center} 
\makebox{\epsfig{file=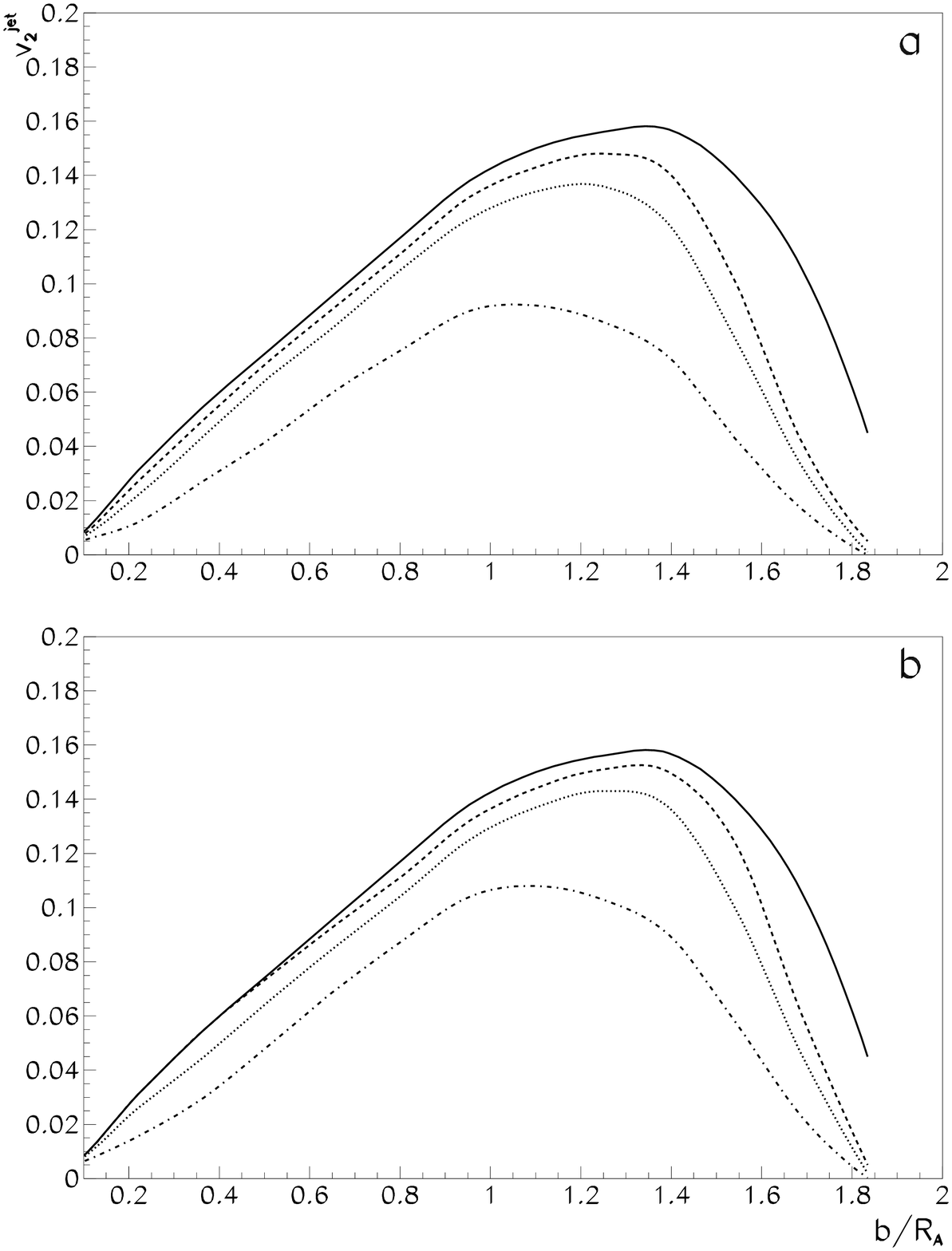, height=200mm}}   
\caption{The impact parameter dependence of ``theoretical'' value of $v^{jet}_2$ 
including collisional and radiative energy loss (solid curve), and $v^{jet}_2$ 
determined by the method (\ref{u2(E)_event}) for $\bar v^-_{cor}/\bar v_2^2 =0$ (dashed 
curve), $0.01$ (dotted curve) and $0.1$ (dash-dotted curve). The result obtained using 
the fourth order cumulant method (\ref{v2E(4)}) coincides with the dashed curve. 
$dN^{\pm}/dy (y=0, b=0) = 3000$ (a) and $6000$ (b).}  
\end{center}
\end{figure}

\end{document}